\documentclass[%
 reprint,
 amsmath,amssymb,
 aps,
]{revtex4-2}

\usepackage{graphicx}
\usepackage{dcolumn}
\usepackage{bm}

\begin{document}

\preprint{APS/123-QED}

\title{Pump depletion in parametric down-conversion with low pump energies}

\author{Jefferson Fl\'orez}
 \email{jflor020@uottawa.ca}
\author{Jeff S. Lundeen}%
\affiliation{%
 Department of Physics and Centre for Research in Photonics, University of Ottawa, 25 Templeton Street, Ottawa ON, K1N 6N5, Canada
}%
\affiliation{%
 Max Planck-University of Ottawa Centre for Extreme and Quantum Photonics, University of Ottawa, Ottawa ON, Canada}

\author{Maria V. Chekhova}
\affiliation{%
 Max Planck Institute for the Science of Light, Staudtstra{\ss}e 2, 91058 Erlangen, Germany}
\affiliation{%
 Institute of Optics, Information and Photonics, University of Erlangen-N\"urnberg, Staudtstra{\ss}e 7/B2, 91058 Erlangen, Germany}
\affiliation{%
Department of Physics, M. V. Lomonosov Moscow State University, Leninskie Gory, 119991 Moscow, Russia}

\date{\today}

\begin{abstract}
We report the efficient generation of high-gain parametric down-conversion, including pump depletion, with pump powers as low as 100 $\mu$W (energies $0.1$~$\mu$J/pulse) and conversion efficiencies up to 33\%. In our simple configuration, the pump beam is tightly focused into a bulk periodically poled lithium niobate crystal placed in free space. We also observe a change in the photon number statistics for both the pump and down-converted beams as the pump power increases to reach the depleted pump regime. The experimental results are a clear signature of the interplay between the pump and the down-converted beams in highly efficient parametric down-conversion sources.
\end{abstract}

\maketitle


Parametric down-conversion (also known as optical parametric generation or optical parametric amplification) is a widely used optical process to generate photon pairs from the interaction of a pump beam with a nonlinear dielectric medium. In high-gain parametric down-conversion (PDC), achieved with pulsed pump beams and tailored nonlinear media, the process is efficient enough to produce bright beams visible to the naked eye, also known as twin beams. More importantly, these twin beams exhibit strong photon number correlations that have no classical counterpart~\cite{Heidmann87,Aytur90,Smithey92,Jedrkiewicz04,Bondani07,Iskhakov09,Brida10a,Harder16}. Thanks to such non-classical correlations, twin beams generated by high-gain PDC find current and potential applications in fields like quantum imaging~\cite{Brida10b,Lopaeva13} and quantum metrology~\cite{Brida10a,Lemieux19}.

PDC has been theoretically studied in the so-called parametric approximation. In such, the pump is a classical and undepleted electric field, while the signal and idler are quantized fields. With this approximation, and in the simplest case of each field being single mode, the mean number $N_\mathrm{PDC}$ of PDC photons scales exponentially with pump field amplitude according to the expression $N_\mathrm{PDC}=\sinh^2(\Gamma t)$. Here $t$ is the interaction time of the pump beam with the medium, while $\Gamma\propto |E_p|d_\mathrm{eff}L$, with $|E_p|$ the pump field amplitude, $d_\mathrm{eff}$ the effective nonlinear susceptibility, and $L$ the length of the medium. It is also known that for a single mode in each of the twin beams, the photon number has a thermal statistical distribution~\cite{Iskhakov12a}. Similar results can be obtained in the more complex case of multi-mode PDC, where the twin beams are generated in many spatiospectral modes.

High-gain PDC is produced in second-order nonlinear optical materials arranged in a variety of ways, such as waveguides, optical cavities, or in free space with single-domain or periodically-poled crystals. The single pass configurations in free space are perhaps the simplest ones in experimental terms given the large amount of commercially available nonlinear crystals. Moreover, they do not require the coupling of light into waveguides, or the alignment of optical cavities or interferometer-like setups in the case of multiple-pass sources. This practical advantage comes at the expense of producing multi-mode PDC and the need for strong pumping in order to reach the high-gain regime. For instance, the required pump power in the case of beta-barium borate (BBO) crystals is of the order of tens of milliwatts. In the case of periodically-poled crystals like potassium titanyl phosphate (PPKTP) or lithium niobate (PPLN), the pump power required for high-gain PDC can be lower due to the use of the strong component of the second-order nonlinear tensor $d_{33}$.

At high-gain, the efficiency of the down-conversion can be so large that the parametric approximation does not hold any more and the evolution of the pump beam should be considered. In particular, the pump may become depleted, substantially converting to signal and idler. At the same time, despite the wide interest on PDC nowadays and the large amount of theoretical and experimental works on this topic, the practical investigation of PDC with an evolving pump field remains elusive, with a few exceptions \cite{Allevi14a,Allevi14b,Perina16}. The novel aspects of PDC with an evolving pump are numerous. From a fundamental point of view, it provides an interesting example of a tripartite system with potential quantum features \cite{Cassemiro07}. In the field of quantum state engineering, it leads to the creation of non-Gaussian states. Furthermore, it might give insights to understand analogous phenomena, as it has been proposed in connection with black holes \cite{Nation10,Bradler16}. On the theoretical side, it remains a challenge to fully describe multi-mode high-gain PDC with pump depletion in a fully quantum mechanical fashion.

In this Letter we report high-gain PDC with pump depletion at pump powers as low as 100~$\mu$W (equivalent to $0.1$~$\mu$J/pulse), and with up to 33\% of the pump energy converted into PDC.
Note that the pump power required to get pump depletion in our PDC source is significantly lower than in BBO crystals~\cite{Allevi14a,Allevi14b,Perina16}. With such low pump powers, we are able to directly compare the pump and PDC photon properties on equal footing up to pump powers at which pump depletion begins. In fact, we observe a complementary behavior in the photon number statistics between the pump and one of the twin beams in the depleted pump regime. We also find a linear growth in the number of PDC photons in the depleted regime, in contrast to the exponential growth when the pump is undepleted, accompanied by a steady behavior for the number of pump photons.

\begin{figure}[htbp]
\centering
\includegraphics[width=\linewidth]{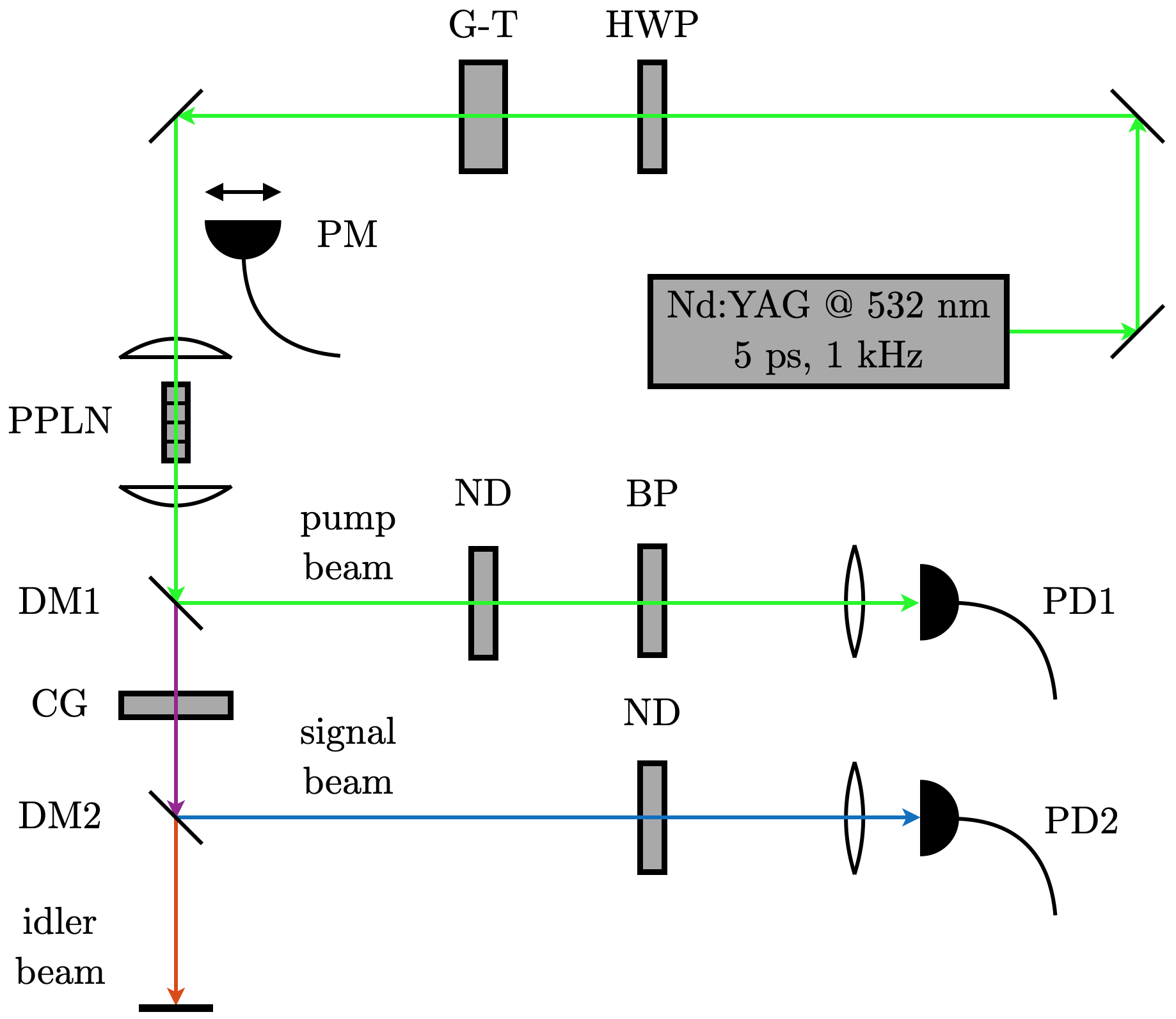}
\caption{Experimental setup. HWP, half-wave plate; G-T, Glan-Thompson prism; PM, power meter; PPLN, periodically-poled lithium niobate crystal; DM, dichroic mirror; ND, neutral density filter; BP, band-pass filter; PD, photodetector; CG, color glass filter.}
\label{fig:ExpSetup}
\end{figure}

As shown in Fig.~\ref{fig:ExpSetup}, we start with the second harmonic of a Nd:YAG laser at 532~nm as the pump beam (EKSPLA PL2210A-1k, 1~kHz repetition rate, 5~ps pulse coherence time). We vary the pump power input to the PPLN crystal via a half-wave plate (HWP) and a Glan-Thompson (G-T) prism. We calibrate the pump power by measuring the maximum transmitted power with a power meter (PM, Thorlabs PM100D, sensor S130VC) and then scaling it according to the theoretical transmission of a rotating HWP plus a fixed polarizer. The light transmitted by the G-T prism is extraordinary polarized in the PPLN crystal to produce collinear type-0 PDC.

We optimize the PDC process to observe pump depletion by tightly focusing the pump beam to a beam waist ($1/e^2$ intensity half-width) of 17~$\mu$m. Our PDC source also benefits from the collinear configuration, where twin beams generated in a certain section of the crystal stimulate the down-conversion process at later sections in a cascade effect. We place the center of the PPLN crystal (5~mm length, 7.9~$\mu$m grating period, fabricated by G\&H~\cite{GandH}) at the waist of the pump beam. After the nonlinear crystal, we collimate the pump beam and separate it from the PDC beams by means of a dichroic mirror (DM1, EKSMA FS mirror, high reflective $>99\%$ for p-polarized light at 532~nm, angle of incidence 45$^{\circ}$). We record the pump power with a photodetector (PD1) after a combination of absorptive neutral density (ND, Thorlabs NE50A) and band-pass (BP, Thorlabs FL532-10) filters. In the PDC path a long-pass colored glass (CG) filter (Thorlabs FGL570) blocks any pump leak before the two PDC beams are separated by a second dichroic mirror (DM2, Semrock FF980). Given the exponential growth in the number of PDC photons, we change the attenuation in the signal beam as function of the input pump power by choosing from a set of ND filters (Thorlabs NE50A, NE20A, NE10A, NE01A). The signal beam centered at 750 nm is then detected by a second photodetector (PD2). The idler PDC beam at 1840 nm is not measured.

Our photodetectors have been implemented and described in previous experimental works~\cite{Hansen01,Iskhakov09,Manceau17}. They are based on a Si PIN photodiode (Hamamatsu S3072) with the same (up to 0.2\%) quantum efficiency (86\%) at both 532 nm and 750 nm. The photodiode is followed by pulsed charge-sensitive amplifiers (Amptek A250 and A275) that transform a detected light pulse into a voltage pulse with certain shape. The area of such a pulse is proportional to the number of photons in the light pulse, so we use an analog-to-digital card to integrate the voltage pulse over time. We calibrate the photodetectors by following a similar procedure to the one already described to calibrate the pump power illuminating the PPLN crystal. The proportionality constants are $6.65(1)\times10^{-12}$ V$\cdot$s/photon and $9.47(3)\times10^{-12}$ V$\cdot$s/photon for PD1 and PD2, respectively. In the calibration process of PD1 we include the BP filter at 532 nm. We report all calibrations and number of photons per pulse in this Letter after averaging 2000 detected light pulses for each input pump power.

\begin{figure}[htbp]
\centering
\includegraphics[width=\linewidth]{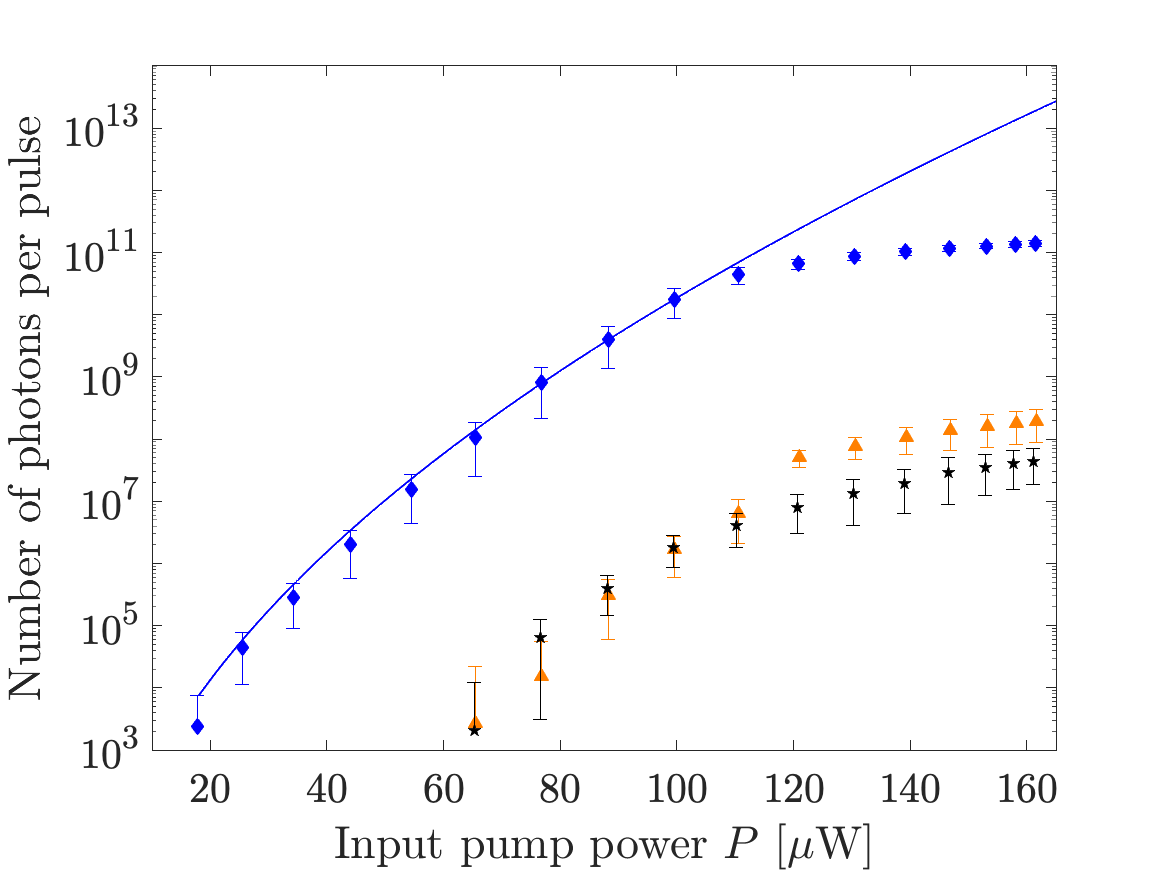}
\caption{Number of photons in logarithmic scale resulting from three different nonlinear processes, namely parametric down-conversion (number of signal photons, blue diamonds), second-harmonic generation (orange triangles) and sum-frequency generation (black stars). The solid blue line corresponds to an exponential fit to the parametric down-conversion data. All experimental error bars in this Letter correspond to three standard deviations for visualization reasons.}
\label{fig:750}
\end{figure}

We present in Fig.~\ref{fig:750} the mean number of detected PDC photons $N_\mathrm{PDC}$ per pulse in the signal beam (blue dimonds) as a function of the input pump power $P$. We observe that $N_\mathrm{PDC}$ displays the exponential growth described by $N_\mathrm{PDC}(P)\propto \sinh^2(b\sqrt{P})$ (solid line) for pump powers up to $P'\approx100$ $\mu$W ($=0.1$~$\mu$J/pulse). Here $b$ is a fitting parameter such that the parametric gain $G=b\sqrt{P'}$ is equal to $12.8(2)$. For $P\gtrsim P'$ the number of signal photons goes from growing exponentially to growing linearly as a result of pump depletion. More specifically the finite energy available in the pump for down-conversion prevents an indefinite exponential growth for $N_\mathrm{PDC}$. The pump powers satisfying $P\gtrsim P'$ define the depleted pump regime. Similar results have already been reported in BBO crystals with pump powers around $\sim 40$ mW ($=80$ $\mu$J/pulse)~\cite{Allevi14a,Allevi14b,Perina16}. In particular, Ref.~\cite{Perina16} also provides a semiclassical model that predicts the exponential and linear behavior for $N_\mathrm{PDC}$ in the undepleted and depleted pump regimes, respectively. We attribute the difference of almost three orders of magnitude in the pulse energies to achieve pump depletion to the focusing of the pump beam and the high optical second-order nonlinearity $d_{33}$ of the lithium niobate crystal, available through quasi-phasematching.

When optimizing our PDC soure we expect to optimize other nonlinear process in the PPLN crystal as well. Even though they are not phase matched, we observe in Fig.~\ref{fig:750} two examples of such processes, namely second-harmonic generation (SHG) at 375 nm due to the frequency doubling of the signal PDC beam at 750 nm (orange triangles), and sum-frequency generation (SFG) at 413 nm due to the interaction of the pump beam at 532 nm and the idler PDC beam at 1840 nm (black stars). According to Fig.~\ref{fig:750}, these nonlinear processes display photon numbers that are several orders of magnitude smaller than PDC. Therefore, we regard them as parasitic processes with a negligible effect on PDC. We also observe in Fig.~\ref{fig:750} that the number of SHG and SFG photons follow a similar trend to the signal PDC ones, i.e. an exponential growth as a function of the input pump power $P$ followed by a linear increase in the depleted pump regime. This trend happens because the nonlinear processes studied here take the PDC photons as input, and therefore must follow the same behavior.

\begin{figure}[htbp]
\centering
\includegraphics[width=\linewidth]{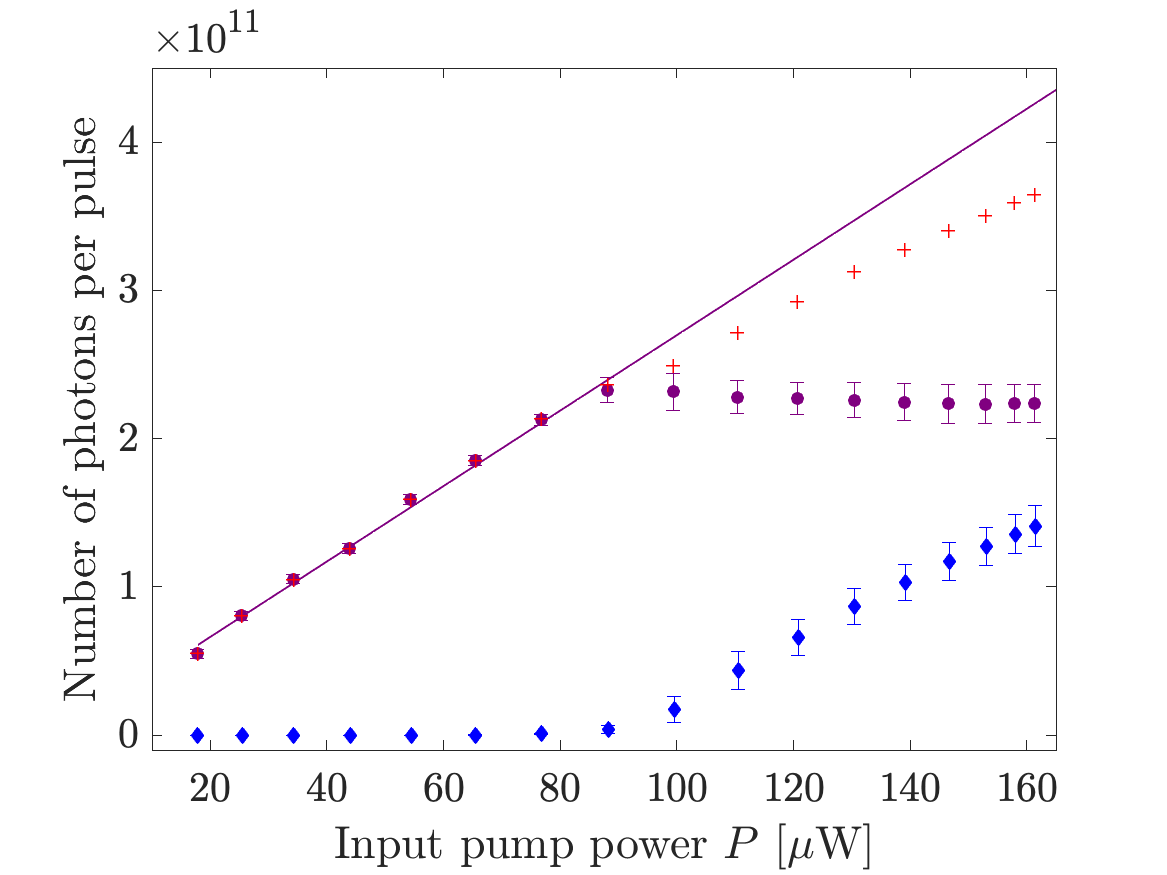}
\caption{Number of signal (blue diamonds) and pump (purple circles) photons after the PPLN crystal, together with their sum (red plus signs). The solid purple line is a linear fit to the number of pump photons in the undepleted pump regime ($P<100$ $\mu$W). The number of SHG and SFG photons presented in Fig.~\ref{fig:750} are not noticeable in linear scale, and therefore we do not present them here.}
\label{fig:532}
\end{figure}

In Fig.~\ref{fig:532} we present the mean number of detected pump photons after the PPLN crystal (purple circles) along with the signal photons (blue diamonds) in a linear scale for comparison. In the undepleted pump regime ($P<P'$) we observe that the number of pump photons grows linearly with $P$, as expected from the linearity of the photodetectors \cite{Manceau17} and losses (15\% reflection at each face of the crystal). In contrast, this number reaches a steady behavior in the depleted regime ($P\gtrsim P'$), and no more photons appear in the pump beam after the crystal regardless of the input pump power $P$. We discard other nonlinear processes as the source of this steady behavior for the pump since they are negligible, as discussed in connection to Fig.~\ref{fig:750}. For $P\approx 160$ $\mu$W, the number of signal photons ($\sim 1.4\times10^{11}$) is 33\% of the expected number of pump ones ($\sim 4.3\times10^{11}$). The latter number is obtained by extrapolating the linear behavior of the number of pump photons into the depleted regime (purple solid line).

To answer the question about where the pump photons go to in the depleted regime, we investigate the overall photon number budget. Specifically, we sum the number of pump and signal photons (red `plus' signs) in Fig.~\ref{fig:532}. Since a pump photon down-converts into a single signal (and idler) photon, this sum must overlap with the linear trend for the pump (purple solid line) for all input pump powers. In the undepleted regime, this condition is fulfilled thanks to the negligible number of PDC photons compared to the pump ones. In the depleted regime we observe a slight discrepancy between the linear trend and the mentioned sum. Such a discrepancy can be explained by unaccounted losses in the ND filters or dichroic mirrors. However, despite this discrepancy, the sum of pump and PDC photon numbers closely follows the linear trend of the pump in the undepleted regime. This result suggests that almost all the input pump photons are converted into PDC.

\begin{figure}[htbp]
\centering
\includegraphics[width=\linewidth]{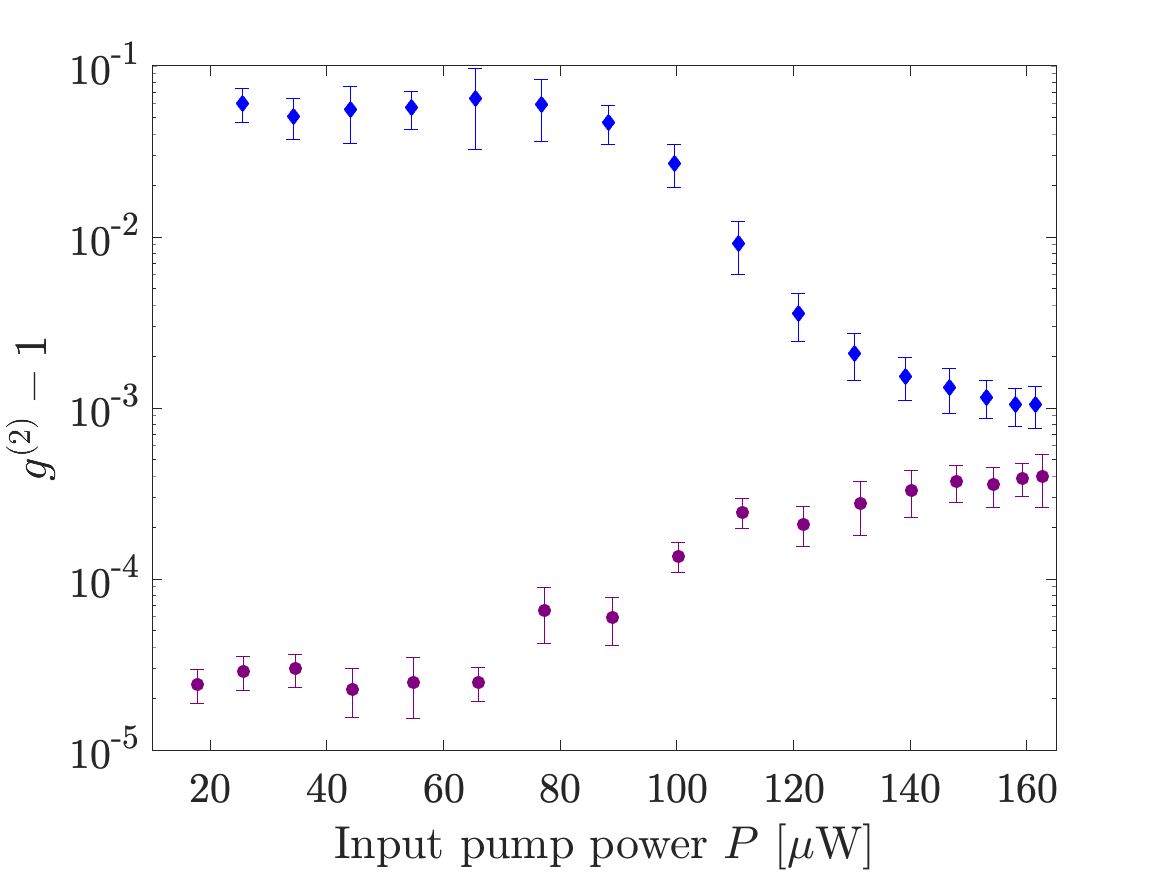}
\caption{Second-order correlation function (minus one) for the pump after the crystal (purple circles) and signal (blue diamonds) beams.}
\label{fig:g2}
\end{figure}

Further, we investigate the photon number statistics of the pump and signal beams by measuring the second-order correlation function at zero time delay, i.e. the bunching parameter $g^{(2)}\equiv g^{(2)}(0)=\langle:\hat{N}^2:\rangle/\langle\hat{N}\rangle^2$. In this expression $\hat{N}$ denotes the photon number operator and $:\ :$ indicates normal ordering, i.e. all annihilation operators are to the right of all creation operators in the product. For large numbers of photons, as in our case, one can neglect the normal ordering and calculate the mean value of the number of photons squared in the numerator of $g^{(2)}$. The bunching parameter is useful to quantify the photon number variance via the expression $\mathrm{Var}(\hat{N})=\langle\hat{N}\rangle+[g^{(2)}-1]\langle\hat{N}\rangle^2$. We present $g^{(2)}$ for the pump after the crystal (purple circles) and the signal (blue diamonds) beams in Fig.~\ref{fig:g2}. In the undepleted pump regime we observe that the quantity $g^{(2)}-1$ is approximately constant for the pump after the crystal and for the signal beam. For the latter, we can use the value of $g^{(2)}$ to estimate the number $M$ of spatiotemporal modes using the expression $g^{(2)}=1+[g^{(2)}_1-1]/M$, with $g^{(2)}_1=2$ for single-mode thermal light \cite{Ivanova06}. As a result, we obtain $M=18(2)$. In contrast, in the depleted regime, $g^{(2)}-1$ displays complementary behaviors for the signal and pump beams. In particular, while $g^{(2)}-1$ decreases around two orders of magnitude for the signal beam, it increases around one order of magnitude for the pump beam.

We can interpret this change of $g^{(2)}-1$ for the signal beam as a deviation from the 18-mode multithermal photon number statistics describing each of the PDC beams in the undepleted regime. The specific reduction of $g^{(2)}-1$ for the signal beam suggests that this beam gets less chaotic in the depleted regime, approaching the statistics of a coherent beam. For the pump beam the interpretation is opposite: the pump beam deviates from the coherent state towards a more chaotic state. We attribute this complementary behavior in the photon-number statistics to a back action of the down-converted fields on the pump, so that the thermal statistics of the PDC beams is imprinted into the pump one. Indeed, when combining the results for $g^{(2)}$ in Fig.~\ref{fig:g2} and the photon-number behaviors in Fig.~\ref{fig:532}, we find a close similarity between our twin-beam source and a laser cavity or an optical parametric oscillator above the threshold. These devices use an active medium or a nonlinear crystal inside an optical cavity to generate coherent light, respectively. According to our results, the statistics of a PDC source optimized to get pump depletion might approach the ones of a source of coherent light without an optical cavity. This can be very useful for  quantum imaging and metrology with twin beams because their thermal noise is a source of additional uncertainties, especially at high brightness.

In summary, by tightly focusing a pulsed pump beam into a PPLN crystal we achieved a PDC conversion efficiency of 33\% accompanied by pump depletion. In particular, we observed that the number of signal photons does not grow exponentially but linearly in the depleted pump regime, with less fluctuations in its photon statistics than in the undepleted case. The excess fluctuations are transferred to the pump, whose number of photons stays nearly constant in the depleted regime. These results were obtained at very low input pump powers, which allows the measurement of the pump beam with the same sensitive photodetectors as the down-converted fields. Given that the pump and PDC fields display similar optical intensities, we expect that our experimental results motivate research towards the interplay of the three fields using a simple experimental configuration.

\begin{acknowledgments}
Canada Research Chairs (CRC) Program, Natural Sciences and Engineering Research Council (NSERC), Canada First Research Excellence Fund (CFREF) award on Transformative Quantum Technologies, Mitacs Globalink Research Award Abroad (IT14583), Colciencias. The authors thank C. Okoth for fruitful discussions.
\end{acknowledgments}


\bibliography{pumpdepletion}

\end{document}